\documentclass[sigconf]{acmart}

\usepackage{comment}
\usepackage{amssymb}
\usepackage{graphicx}
\usepackage{amsmath}
\usepackage{url}
\usepackage{float}
\usepackage{lscape}
\usepackage{esvect}
\usepackage[linesnumbered,ruled]{algorithm2e}
\usepackage{algpseudocode}
\usepackage{footnote}
\usepackage{enumitem}
\usepackage{tablefootnote}

\AtBeginDocument{%
  \providecommand\BibTeX{{%
    \normalfont B\kern-0.5em{\scshape i\kern-0.25em b}\kern-0.8em\TeX}}}

\copyrightyear{2020}
\acmYear{2020}
\setcopyright{acmcopyright}\acmConference[ICTIR '20]{Proceedings of the 2020 ACM SIGIR International Conference on the Theory of Information Retrieval}{September 14--17, 2020}{Virtual Event, Norway}
\acmBooktitle{Proceedings of the 2020 ACM SIGIR International Conference on the Theory of Information Retrieval (ICTIR '20), September 14--17, 2020, Virtual Event, Norway}
\acmPrice{15.00}
\acmDOI{10.1145/3409256.3409837}
\acmISBN{978-1-4503-8067-6/20/09}

\begin{document}
\fancyhead{}


\title{Efficient Test Collection Construction via Active Learning}

\author{Md Mustafizur Rahman}
\affiliation{%
  \institution{The University of Texas at Austin}
  \streetaddress{}
}
\email{nahid@utexas.edu}

\author{Mucahid Kutlu}
\affiliation{%
  \institution{TOBB  University of Economics and Technology}
  \streetaddress{}
}
\email{m.kutlu@etu.edu.tr}

\author{Tamer Elsayed}
\affiliation{%
  \institution{Qatar University}
  \streetaddress{}
}
\email{telsayed@qu.edu.qa}

\author{Matthew Lease}
\affiliation{%
  \institution{The University of Texas at Austin}
  \streetaddress{}
}
\email{ml@utexas.edu}

\renewcommand{\shortauthors}{Rahman, et al.}
\begin{abstract}
To create a new IR test collection at low cost, it is valuable to carefully select which documents merit human relevance judgments.  Shared task campaigns such as NIST TREC pool document rankings from many participating systems (and often interactive runs as well) in order to identify the most likely relevant documents for human judging. However, if one's primary goal is merely to build a test collection, it would be useful to be able to do so without needing to run an entire shared task. Toward this end, we investigate multiple active learning strategies which, without reliance on system rankings: 1) select which documents human assessors should judge; and 2) automatically classify the relevance of additional unjudged documents. To assess our approach, we report experiments on five TREC collections with varying scarcity of relevant documents. We report labeling accuracy achieved, as well as rank correlation when evaluating participant systems based upon these labels vs.\ full pool judgments. Results show the effectiveness of our approach, and we further analyze how varying relevance scarcity across collections impacts our findings. To support reproducibility and follow-on work, we have shared our code online\footnote{\url{https://github.com/mdmustafizurrahman/ICTIR_AL_TestCollection_2020/}}.
\end{abstract}


\maketitle

\section{Introduction}
Test collections provide the foundation for Cranfield-based evaluation of information retrieval (IR) systems~\citep{sanderson2010test}. Unfortunately, it has become increasingly expensive to manually judge so many documents as collection sizes have grown. On the other hand, failing to collect sufficient relevance judgments can compromise evaluation reliability. Even commercial search engines, despite their query logs, still rely on large teams of human assessors \citep{google-guidelines-2016}. Consequently, there is great interest in developing more scalable yet reliable IR evaluation methodology. 

To create a new IR test collection at minimal cost, it is valuable to identify a minimal set of documents for human relevance judging. This is typically accomplished by running a shared task campaign, such as NIST TREC, then pooling search results from many participating systems (and often interactive runs) to identify the most likely relevant documents for  judging \citep{cormack1998efficient}. While this approach is now canonized in IR practice, organizing the community to run a shared task is complicated, slow, and requires many hours of work by organizers and participants. This hidden, real-world cost may far exceed simple judging costs, which are often the only measure of cost reported. This suggests a more complete accounting of cost ought to be considered, if not quantified, wrt.\ building IR test collections. Shared tasks have many other benefits, but if one's primary goal is merely to build a new test collection, it would be useful if this could be achieved without needing to run a shared task \citep{sanderson_forming_2004}. 

In this paper, we investigate the following research question: {\em how feasible is it to build a new test collection without a shared task, and how can one best accomplish this?} To this end, we explore active learning (AL) \citep{settles2012active} methods to support test collection construction without reliance on shared task document rankings. Rather than develop novel active learning algorithms, our focus in this work is the novel combination of active  learning methods and inferred assessments for building test collections without running a shared task. To the best of our knowledge, this has not previously been pursued in the literature. 

Our approach involves learning a topic-specific document classification model for each search topic. We consider two distinct applications of AL. Firstly, we apply AL to select which documents assessors should judge, and we explore two document selection strategies \citep{cormack2014evaluation}: {\it continuous active learning} (CAL) and {\it simple active learning} (SAL). Secondly, we consider use of AL to  automatically classify relevance of additional unjudged documents. This differs from traditional IR evaluation, which often ignores unjudged documents or assumes them to be non-relevant. Moreover, the ability to use any hybrid combination of human and automatic judgments in evaluation provides a flexible tradeoff space for balancing cost vs.\ accuracy  \citep{nguyen2015combining}. Though others have pursued automatic or semi-automatic relevance labeling \citep{carterette2007semiautomatic,buttcher2007reliable,hui2015selective, cormack2005spam, grossman2016trec}, prior studies do not use AL for i) selecting documents for annotations and ii) inferring relevance labels for unjudged documents simultaneously in constructing IR test collections. 

Because AL is supervised, an initial {\em seed set} of labeled documents is needed to bootstrap learning. We consider two distinct scenarios for how these seed judgments might be obtained: {\em interactive search} (IS) and {\em Rank-based Document Selection} (RDS). We emphasize that these represent alternative scenarios rather than competing methods. IS assumes topic assessors utilize an IS system during a careful topic creation process, as traditionally practiced in TREC. This produces seed judgments as a free by-product. RDS, on the other hand, assumes a scenario like the TREC Million Query Track \citep{carterette2009if} in which topic formation is extremely brief and assessors are not provided an IS system in which to explore the collection. In this scenario, an off-the-shelf IR system is used instead to produce a single document ranking; assessors then judge documents in this rank-order until enough seed judgments have been collected to kickstart AL. 

In exploring our central research question, \textbf{contributions} of our work are as follows: 1) We show that it is feasible to develop IR test collections without needing to organize and run a shared task (i.e., just to identify potentially relevant documents for judging); 2) We demonstrate how AL can be effectively applied to test collection construction via: i) document selection for collecting human relevance judgments;  and ii) automatic labeling of additional unjudged documents; and 3) We investigate three document selection methods and two seed data scenarios across five TREC tracks. For one document collection (TREC TIPSTER disks 4-5), we present the first work we know of going beyond the pool to automatically judge the rest of the document collection.

\section{RELATED WORK}
\label{section:related-work}
Ever-larger document collections challenge systems-based Cranfield \citep{cleverdon1967cranfield} evaluation of IR systems due to needing to collect so many relevance judgments. 
While many methods now exist to intelligently select which documents to judge, these methods typically assume a shared task context (e.g., TREC) in which document rankings from  many participating systems are available. In contrast, we want to be able to construct a new test collection {\em without} needing to run a shared task \citep{soboroff2013building, sanderson_forming_2004}.  


\citet{buttcher2007reliable} propose labeling unjudged documents using a classifier trained on a subset of pool documents. This subset of pool documents is developed by considering documents ranked by a subset of the submitted rank systems. The trained classifier is then used to predict   relevance labels of documents which are ranked by the remaining set of rank systems in the shared task. 
We both report results for the same 2006 Terabyte Track run, but our results are not directly comparable to theirs because they assume a traditional machine learning setup, whereas we  motivate and adopt the {\em finite-pool} evaluation setting proposed in \citep{cormack2014evaluation}. However, we effectively reproduce their method as a baseline, using logistic regression and random  document selection. We show strong improvement over this baseline.

While \citet{hui2015selective} use document rankings information in their own proposed method, they also reproduce \citet{buttcher2007reliable}'s SVM method as a baseline, reporting results on the same WebTrack 2013 and 2014 collections we use in this study. However, as with \citet{buttcher2007reliable}, they do not evaluate their approach under a finite-pool scenario. Though this means that our results are not directly comparable, our same baseline configuration described above roughly reproduces their SVM approach. 

For AL document selection, we evaluate the same CAL and SAL methods \cite{Rahman2019www} that \cite{cormack2014evaluation} assess in the domain of e-discovery, where they focus on set-based rather than ranked retrieval. Moreover, judging cost is also measured differently in e-discovery: no document can be ``screened in'' automatically since all must be reviewed for privilege following discovery. 
Recently, \citet{cormack2018beyond} propose a variant of ``S-CAL'' \citep{cormack_scalability_2016}, which rather than selecting the highest-scoring documents for relevance judgment, randomly samples some documents from those the highest-scoring  documents for annotation. They report results on the collected human relevance judgments (e.g. TREC pool documents) but not hybrid judging. 

\citet{carterette2007semiautomatic} apply \citet{carterette2006minimal}'s document selection method to iteratively collect relevance judgments. Based on the cluster hypothesis, their per-topic logistic regression classifier estimates the probability of relevance of an unjudged document conditional on its similarity to other judged documents in the cluster (e.g. relevant and non-relevant document clusters). A key difference with our work is that their document selection strategy relies on having run a shared task. 


Similarly, \citet{nguyen2015combining} investigate AL-based 
relevance judging in the domain of systematic-review in medicine, which bears much in common with e-discovery \citep{Lease16-medir}. As above, AL is used to reduce labeling costs, but without intent to construct a test collection or evaluate IR systems based on automatic labels. They also adopt a finite-pool evaluation setting, but unlike us, they use both trusted judges and crowds in combination for human judging.


\citet{rajput2012constructing} develop a framework for constructing a test collection using an iterative process between updating nuggets and annotating documents. 
However, because their automatic nugget extraction fails to extract nuggets from documents which are difficult to parse (e.g. TREC Web Track), the authors fall back to using document rankings from participating systems of a shared task evaluation. \citet{li2017active} also utilize a shared task by inducing a probability distribution from the participating systems and a probability distribution over the ranks of the documents. They then actively sample documents from the joint distribution to construct an unbiased test collection. 

\section{PROPOSED APPROACH}
\label{section:method}



\subsection{Task Definition and Learning Model}
\label{section:defintion}

We assume the Cranfield model of system-based IR evaluation that is based on pre-defined search topics and relevance judgments. In order to construct a hybrid human-machine system for binary relevance judging of collection documents, we induce a topic-specific binary classifier $c^j$ for each search topic $j$ in the topic set $T$ of $n$ topics. 
Assume we have a document collection $X$ of $m$ documents 
(represented by extracted features). Let $y^i_j$ denote the binary relevance judgment for <document $i$, topic $j$>. The training data for topic $j$ is comprised of a set of pairs $\langle x^i, y^i_j \rangle$.

For each search topic for which we wish to train a topic-specific classifier $c^j$, we must collect topic-specific training data. 
As we utilize the probability of relevance $p(y^i_j|x^i)$ in the document selection criteria, we adopt logistic regression\footnote{\label{scikit} We utilize the implementation of logistic regression and support vector machine (SVM) from Scikit-learn package (\url{https://scikit-learn.org/}), a machine learning library in Python programming language.} as our learning model to infer the probability of relevance $p(y^i_j|x^i)$ for each document $x^i$ for topic $j$:



\begin{equation}
p(y^i_j|x^i) = h_\theta(x_i) = \frac{1}{1 + \exp(-\vv{\theta}^Tx_i)} 
\end{equation}
with $\vv{\theta} \in \mathbb{R}^D $ model parameters. 
We set $\lambda = 10^{-8}$ in all reported experiments after tuning on \cite{maas-EtAl:2011:ACL-HLT2011}'s dataset.  We adopt the canonical TF-IDF 
representation of documents.

In  unreported experiments  (due to lack of space), we compared logistic regression vs.\ support vector machine (SVM)\textsuperscript{\ref{scikit}} and XGBoost\footnote{ We use the implementation of XGBoost from \url{https://xgboost.readthedocs.io/en/latest/python/python_intro.html}} models. Evaluating F1 accuracy on \cite{maas-EtAl:2011:ACL-HLT2011}'s dataset, we found that tuned logistic regression performed comparably to the tuned SVM and marginally better than XGBoost. Moreover, in terms of wall clock time, logistic regression is far faster than SVM and XGBoost, which is a significant advantage for AL. 




    
    

\begin{algorithm}
    \SetKwInOut{Input}{Input}
    \SetKwInOut{Output}{Output}
    \Input{~~Document collection $X$ $\bullet$ batch size $u$ $\bullet$ total budget $b$}
    \Output{~~Relevance judgments $R^{1:n}$ for topics $1:n$}

%
    
    \For{\textnormal{topic} $j\gets1$ \KwTo $n$} {
    	
        Select seed documents $S\in X$ for topic $j$ 
    	\\
        $R^j \leftarrow \{\langle x^i, y^i_j \rangle\ |\ x^i \in S\}$ \Comment Collect initial judgments \\
    	Learn relevance classifier $c^j$ using $R^j$\\
    	$b \leftarrow b - |S|$ \Comment Update remaining budget 
    }
    \While {True}{
    \For{\textnormal{topic} $j\gets1$ \KwTo $n$}{
    	\algorithmicif\ $b < u$ \algorithmicthen\ \Return \Comment Budget exhausted \\
       	$\forall x\in X$ predict topical relevance of document $x$ using $c^j$\\
        Select $u$ documents $S \in X$ to judge next for topic $j$ 
        \\
        $R^j \leftarrow R^j \cup \{\langle x^i, y^i_j \rangle\ |\ x^i \in S\}$ \Comment Collect judgments\\
       	Re-estimate relevance classifier $c^j$ using expanded $R^j$\\
       	$b \leftarrow b - u$ \Comment Update remaining budget
    }}
    
\caption{Active Learning Algorithm}
\label{algorithm: active learning}
\end{algorithm}

\subsection{Active Learning}
\label{section:active_learning}



An active learning~\citep{settles2012active}  algorithm iteratively selects which document $x^i$  should be labeled next in order to maximize the classifier's learning curve for each topic. This
reduces the amount of human effort required to induce an effective model. {\bf Algorithm \ref{algorithm: active learning}} describes our active learning strategy to develop a test collection. The first loop (Lines 1-6) collects the seed document labels for each topic 
and trains a topic-specific document classifier using the seed documents.  
In the second loop (Lines 7-14), the learned classifier is used to select documents for further annotation. For selecting documents (Line 10), we employ the strategies discussed in the {\em Document Selection Criteria} section below. Those further annotated documents are employed to re-train the topic-specific classifier. This process continues iteratively until we exhaust the allocated budget. 
\subsubsection{Seed Document Selection}
\label{section:method_seed}

In order to learn a topic-specific document relevance classifier, topic-specific training data is needed. We assume that no such labeled data for each topic exists in advance. 
Consequently, we must collect an initial {\em seed set} of human relevance judgments for each search topic in order to initialize our AL model. 
While 
we could simply select a (uniform) random sample, 
it is unlikely with large class imbalance that such random selection would find any relevant documents; just imagine randomly sampling documents from the Web in order to find a relevant document for a particular topic. 
Instead, we consider two scenarios:

{\bf Interactive Search (IS)}: IS is also known as {\em search-guided assessment} \citep{oard2004building}.
%
Our IS scenario assumes traditional TREC practice for constructing search topics. 
The assessor searches the document collection in order to find some minimal set of relevant and non-relevant documents in order to establish the topic as viable. If insufficient relevant documents are found, the topic is discarded (topics with too few relevant or non-relevant documents 
provide little information for IR system evaluation). 
Whereas the NIST process above would traditionally discard these initial judgments, we would instead keep them as the seed set for active learning.  As such, we would essentially get seed documents for AL for free as a by-product of topic development, but we nevertheless include the cost of these judgments like any others in our reported experiments.  



{\bf Rank-based Document Selection (RDS)}: 
This scenario supports seed data selection when there is no interactive search interface or when no judgments are collected during topic formation, as in the TREC Million Query (MQ) Track \citep{carterette2009if}. 
Instead, we assume access to a single, moderately effective off-the-shelf or in-house IR system. Given a search query as input, the document collection is searched to produce a ranked list of documents. The assessor is then asked to proceed down the ranked list until at least $k$ relevant and non-relevant document(s) have been found, or some maximum effort is reached without success. In this latter case, the topic is discarded, as in typical TREC practice discussed in the IS scenario above. While shared task pooling identifies top-ranked documents for judging, 
RDS selects seed documents via only a single system ranking, then relies on AL to identify further relevant documents in order to create a robust topic.

\subsubsection{Batch Active Learning}
We assume AL selects the next most informative instance to label from a fixed set of unlabeled instances --- 
the set of unlabeled documents \citep{lewis1994sequential}. 
We also assume {\em batch learning}, in which at each time step, we select $u$ unlabeled examples to label next. 


\subsubsection{Document Selection Criteria}
\label{subsubsection:documentselection}
We consider three document selection strategies \citep{cormack2014evaluation}: {\em Simple Passive Learning} (\textbf{SPL}), {\em Simple Active Learning} (\textbf{SAL}),  {\em Continuous Active Learning} (\textbf{CAL}).  

\textit{SPL} selects documents uniformly at random, corresponding to traditional supervised learning in which training data is assumed to be sampled i.i.d.\ from the domain. Including passive SPL provides a useful comparison vs.\ active selection methods. 

\textit{SAL} selects the document to label next for which the current model is maximally uncertain of its correct label, such that labeling this document is expected to maximally inform the current model. We adopt a common uncertainty function based on entropy: 
\begin{equation}
{\text Uncertainty}(x)  = - \sum_{y \in Y} p(y|x) \log p(y|x) 
\end{equation}
where $p(y|x)$ is the a posteriori probability from the classifier and $y$ is relevant or non-relevant. 
With binary relevance, SAL selects: 
\begin{equation}
x^\star = \arg\!\min\limits_{i}\ |p({\text relevant}|x^i)-0.5|
\end{equation}

With \textit{CAL}, the learning algorithm selects the unlabeled document which the current model predicts is most likely to be relevant:
\begin{equation}
x^\star = \arg\!\max\limits_{i}\ p({\text relevant}|x^i)
\end{equation}
While SAL is more commonly used in AL, \citet{cormack2014evaluation} find that CAL is more effective. However, note that their task goal is to find as many relevant documents as possible, assuming assessors must manually label any relevant documents. In contrast, our goal is to optimize a human-model hybrid system. 

\begin{table}
\scriptsize
\begin{center}
\caption{Test collection statistics. As collections have grown larger, judging budgets have also shrunk, leading to increased prevalence of relevant documents in later tracks.}
\begin{tabular}{|l|c|c|r|r|r|r|}
\hline
\bf Track & \!\!\bf Collection\!\! & \bf Topics  & \bf \#Docs & \!\!\bf \#Judged\!\!  & \bf \%Rel \\
\hline
\!\!2014 Web Track (WT'14)\!\! & ClueWeb12 & 251-300 & 52,343,021 & 14,432 & 39.2\%\!\!  \\
\!\!2013 Web Track (WT'13)\!\! & ClueWeb12 & 201-250 & 52,343,021 & 14,474 & 28.7\%\!\!   \\
\!\!2006 Terabyte Track (TB'06)\!\! & Gov2 & 801-850 & 25,205,179 & 31,984  & 18.4\%\!\! \\
\!\!1998 Adhoc Track (Adhoc'98)\!\! & \!\!Disks 4,5\tablefootnote{Adhoc'98 and Adhoc'99 tracks exclude \textit{congressional record} from TIPSTER Disks 4,5}\!\! & 351-400 & 528,155 & 80,345 & 5.8\%\!\!  \\
\!\!1999 Adhoc Track (Adhoc'99) \!\! & \!\!Disks 4,5\!\! & 401-450 & 528,155 & 86,830 & 5.4\%\!\!  \\
\hline
\end{tabular}
\label{table:characteristic_data}
\end{center}
\end{table} 
%

\section{EXPERIMENTAL SETUP}
\label{section:experimental_setup}

\subsection{Datasets and Preprocessing}
\label{section:preprocessing}


We conduct our experiments on five TREC tracks  (see \textbf{Table \ref{table:characteristic_data}}). 
As shown in Table \ref{table:characteristic_data}, later tracks show increasing prevalence of relevant documents in judged pools, from approximately 5\% to almost 40\%. 
Note that since we assume binary relevance in this work, we collapse NIST graded relevance judgments to binary. 


To select seed documents under the IS setting, we assume 5 relevant and 5 non-relevant seed judgments for all topics are produced during topic creation (otherwise the topic would have been discarded and never used). We report the cost of these judgments like any other judgments collected during AL (i.e., cost of 10 here). Over all 5 collections, only 5 total topics were found to have < 5 relevant documents, and so only these 5 topics were discarded (consistently across all reported experiments). 

For the RDS setting, rather than running our own IR system, we randomly select an existing ranking from each TREC track  from the set of participating systems (see {\bf Table \ref{table:MAP_for_rankers}} for statistics of our randomly-selected system vs.\ statistics
of other participating systems). We assume the assessor proceeds
down the ranked list until at least 1 relevant and 1 non-relevant
document is found, after which we proceed using AL. 

\begin{table}[t]
\begin{center}

\caption{MAP scores of systems used for Rank-based Document Selection (RDS) vs.\ track average and std.\ deviation.}

\begin{tabular}{|l|c|c|c|}
\hline
\bf Track & \bf MAP & \bf Track Avg.\ & \!\!\bf Track STD.\ \\
\hline
WT'14 & 0.181 &  0.165 & 0.065 \\
WT'13 &  0.111 & 0.115 & 0.041 \\
TB'06 &  0.350 & 0.276 & 0.089 \\ 
Adhoc'98 & 0.186 & 0.194 & 0.080\\
Adhoc'99 & 0.260 & 0.234 & 0.096 \\
\hline
\end{tabular}
\label{table:MAP_for_rankers}
\end{center}
\end{table} 

\subsection{Pool Document Collection vs. Complete Document Collection}
One question in experimental setup is how to handle documents outside of the TREC judgment pools, for which no human relevance judgment is available. We describe here several alternative evaluation settings for addressing this.  Note that this is only an evaluation issue, and does not alter the actual AL algorithms.

{\bf 1) Pool Document Collection.} In this setting, we restrict AL to the original set of pool documents, meaning that it can only request labels for pool documents. For any such selected document, we simply lookup the original NIST assessor judgment for it (i.e., retrospective AL evaluation based on previously collected labels). We consider two applications of AL in this setting: 
 
{\bf i) Human-only Judging}. 
Traditional practice uses only  human judgments to label pool documents. We first evaluate AL for {\em document selection} only, i.e., determining the best set of documents for human assessors to judge for a given assessment budget in order to best evaluate IR systems. As the size of the evaluation budget approaches the full pool size, we converge to the original TREC results. 

{\bf ii) Hybrid Judging}. Next, we further consider using AL to further automatically label the relevance of remaining pool documents for which the AL system has not requested a human judgment. We then evaluate whether such hybrid judging improves evaluation vs.\ using only human judgments. As the size of the evaluation budget approaches the full pool size, we converge to human-only judging. 

{\bf 2) Complete Document Collection.} In this evaluation setting, we consider applying AL to label the complete document collection, rather than only the pool. 
While NIST relevance judgments exist for pool documents, whenever AL requests a label for a non-pool document, the evaluation framework must somehow still provide a label. As an expedient solution, the evaluation framework simply returns a non-relevant label to AL for all non-pool documents. While this solution is far from perfect, since AP also assumes non-judged documents are non-relevant, this should tend to bias AL predictions in a consistent direction with original TREC pool results based on AP.


\subsection{The {\em Finite-Pool} Scenario for Evaluating Active Learning} 
\label{subsection:finite-pool}


The goal of our task is to maximize labeling accuracy relative to cost (i.e., the number of human judgments requested). At one end of the decision-space, we could use human judging exclusively and forgo automatic classification altogether. This corresponds to traditional human-only relevance judging. Because we assume human assessors are infallible, this maximizes labeling accuracy, but this high accuracy comes at maximal cost. At the other extreme, all documents could be automatically classified. This minimizes cost (since it completely eliminates human judging) but also minimizes accuracy, since there are no human labels collected to train the classifier. Between these two extremes lies a rich decision-space of hybrid judging in which labeling is divided between man and machine \citep{nguyen2015combining}. 

This AL evaluation setting is known as a {\em finite-pool} \citep{nguyen2015combining}, and it differs markedly from typical machine learning (ML) evaluation. Typically, one trains a classifier on one set of data then evaluates it on a separate test set, assessing classifier generalization from training data. In contrast, the finite-pool setting has no prior training data, nor do we care about classifier generalization to some future testing data; all we care about is labeling the present, finite set of documents before us. Following this, any learned classifier is simply discarded. 
We believe this evaluation setting best represents the actual scenario interest --- building a new test collection --- though we diverge here from related prior work \citep{buttcher2007reliable,hui2015selective}. With regard to terminology, note that the term {\em finite-pool} comes from AL and has no relation to {\em pooling} in IR evaluation. This  unfortunate terminology collision stems from bridging these disparate literatures.

\subsection{Evaluation Metrics}

{\bf Effectiveness Curves and AUC}.\ We present our results as plots showing cost vs.\ effectiveness of each method being evaluated at different cost points  (corresponding to varying evaluation budget sizes). In general, different test collections might be built with different annotation budgets. We also report Area Under Curve (AUC) across all cost points, approximated via the Trapezoid rule. 

{\bf Cost}.\ We measure cost with regard to manual judging budget, i.e., the cost of human judgments (assuming automatic classification is free). 
%
We report cost in batch size increments. Specifically, we use 10\% of the pool size for each topic as the batch size, reporting results at \{0\%, 10\%, 20\%, \ldots, 100\%\} human judging of each topic's pool (where 100\% corresponds to the original TREC pool size). Note that we assume cost of each human label as constant, whether it be in seed judging (IS or RDS) or during AL. 
%

{\bf Labeling Accuracy.} We measure our hybrid (human + automatic) AL labeling accuracy in terms of $F_1$, as averaged across topics.



{\bf Rank Correlation}.\ We also assess the reliability of using our labeling methods to evaluate IR systems. A relative performance ranking of participating systems in each track is then induced based on these metrics. As ground truth ranking, we calculate inferred AP (infAP) \citep{yilmaz2006estimating} scores for participating systems using all NIST judgments, as computed via standard {\tt trec\_eval}.
We then compute the Kendall's $\tau$ rank correlation \citep{kendall1938new} between the ground-truth system ranking vs.\ our proposed method's ranking. By convention \citep{voorhees2000variations}, $\tau=0.9$ is assumed to constitute an acceptable correlation level for reliable IR evaluation.

\begin{figure*}[ht]
\centerline{\includegraphics[width=0.75\textwidth]{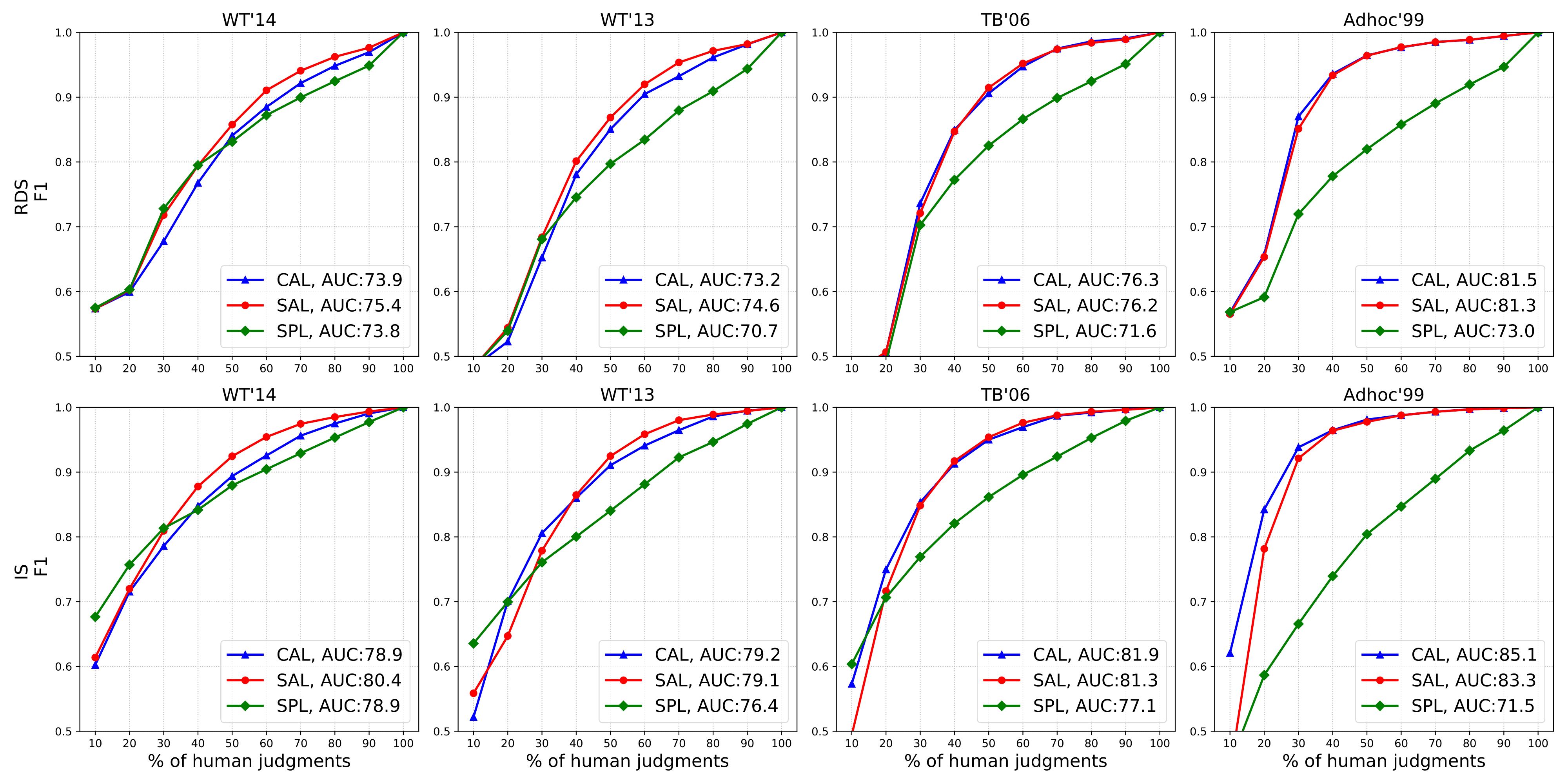}} 
\caption{Human judging cost (x-axis) vs.\ F1 classification accuracy (y-axis) for hybrid human-machine judging of document pools for four TREC Tracks. The \% of human judgments on x-axis is {\em wrt.} the number judged for each collection (Table \protect\ref{table:characteristic_data}).}
\label{Figure:classification}
\end{figure*}

\section{RESULTS AND DISCUSSION}

\subsection{Evaluation on the Pool Document Collection}
In this section, we present experimental results on the Pool Document Collection. We begin by evaluating labeling accuracy of our hybrid AL approaches. This is followed by reporting Kendall's $\tau$ rank correlation results using AL for (i) human-only (incomplete) judging of pool documents; and (ii) hybrid human-machine (complete) labeling of all pool documents. For the Pool Document Collection setting, we omit reporting results on the Adhoc'98 collection due to space constraint.


\subsection{Classification Results on the Pool Document Collection}
\label{subsection:documentclassification}
We first report F1 labeling performance of our hybrid AL approaches. As discussed before, we consider two scenarios for how seed documents are selected to initialize AL: IS and RDS. 

{\bf Figure \ref{Figure:classification}} presents F1 performance results of the three document selection approaches: SPL, SAL, and CAL. 
The x-axis of each plot indicates the percentage of pool documents manually judged (using NIST labels), with the remainder of the pool automatically labeled by the classifier. 
NIST judges are treated as infallible, so all methods ultimately converge to 100\% F1 at the right-end of each plot, corresponding to complete manual judging of the entire pool.

{\bf Seed Selection Scenarios}. As noted earlier, IS and RDS results should be understood as corresponding to distinct scenarios, rather than alternative methods to be compared vs.\ one another.

{\bf Active vs.\ Passive Learning}. Comparing active learning (SAL and CAL) against passive learning (SPL) methods, 
%
Figure \ref{Figure:classification} shows  that SAL or CAL consistently outperforms SPL in terms of AUC. 
We also see that for TB'06 and Adhoc'99 
collections, both CAL and SAL with IS perform comparably, requiring around 30\% (TB'06) and 40\% (Adhoc'99) of human judgments to achieve 90\% F1 measure. In contrast, SPL requires 60\% for TB'06 and 70\% for Adhoc'99. 


{\bf CAL vs. SAL.} It is evident from Figure \ref{Figure:classification} that CAL consistently provides better performance than SAL in terms of AUC in low prevalence test collections (TB'06 and Adhoc'99). On the other hand, for high prevalence test collections (WT'14 and WT'13) SAL dominates CAL (e.g. for 3 out of 4 different plots SAL wins over CAL). 
This finding is consistent with that of \citet{cormack2014evaluation}, despite the various differences between our studies discussed earlier.

{\bf Varying Scarcity of Relevant Documents}.  
Prevalence of relevant documents can vary widely across different  test collections, as well as across topics within a single test collection. For example, WT'14 has the highest average prevalence (around 40\%), while Adhoc'99 has only 5\% average prevalence of relevant documents. Looking at Figure  \ref{Figure:classification}, we see that 
varying prevalence plays an important role in explaining the differing performance of AL vs.\  passive learning. For TB'06 and Adhoc'99, where we have a low prevalence rate (less than 20\%), SAL and CAL outperform SPL by a large margin. 
However, with the higher prevalence rates in WT'13 and WT'14, SPL performs much better, though is still outperformed by SAL. Another notable observation is that as we move from higher prevalence collections (e.g., WT'14) to lower prevalence collections (e.g., Adhoc'99), AUC of CAL with IS consistently increases; the same does not always hold for SAL.




\begin{figure*}[h]
\centerline{\includegraphics[width=0.75\textwidth]{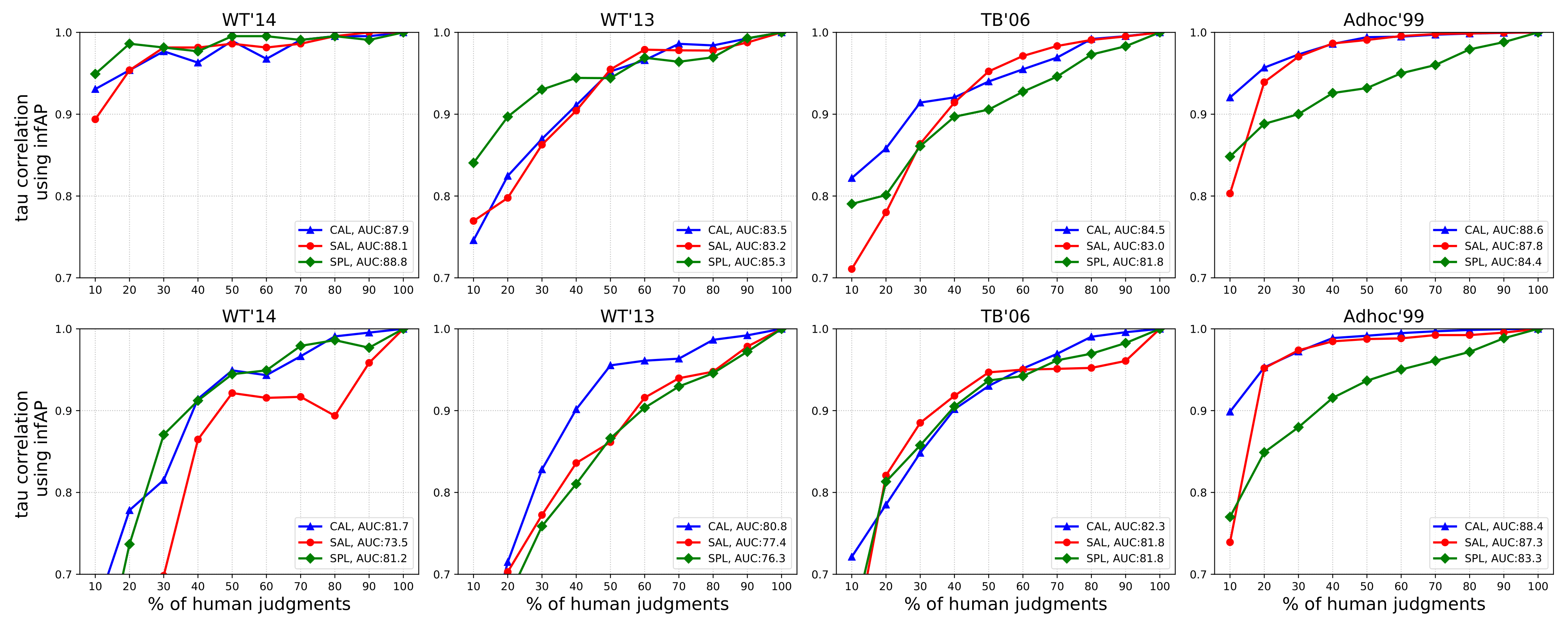}} 
\caption{
Kendall's $\tau$ rank correlation achieved with and without automatic labeling. \underline{Top Row}: hybrid human-classifier labeling. \underline{Bottom Row}: human judgments only. Systems are evaluated using infAP. Ground truth ranking is induced by system infAP scores using all NIST judgments. 
} 
 
\label{Figure:infaptaucorrelation}
\end{figure*}

\subsection{Rank Correlation Results on the Pool Document Collection}
\label{section:rank_correlation}

As discussed before, we consider two applications of AL for aiding IR test collection construction: 1) selecting documents for human judging; and 2) further automatic labeling of all remaining unjudged documents (in the pool). We evaluate these two approaches in turn. Results in this section assume the IS setting.

{\bf Figure \ref{Figure:infaptaucorrelation}} presents Kendall's $\tau$ rank correlation results using infAP. The x-axis indicates the percentage of the pool manually judged, and the y-axis indicates $\tau$ correlation. We plot results for CAL and SAL  strategies, as well as baseline SPL. 

{\bf Using Human-only Judging}. We first consider evaluating IR systems using only human judgments of the documents selected by AL. This is shown in the bottom row of the figure. We see that CAL consistently achieves the highest $\tau$ correlation wrt.\ AUC for  infAP. 
%



{\bf Using Hybrid Judging}. We next consider the second condition of hybrid judging: automatically labeling the remainder of pool documents beyond those manually judged.  This is shown in the top row of Figure \ref{Figure:infaptaucorrelation}. Again we can see that prevalence ratio plays an important role. For example, for low prevalence collections (e.g.\  TB'06 and Adhoc'99), AL with hybrid judging far exceeds performance of SPL. In contrast, for high prevalence collections, SPL surprisingly outperforms CAL and SAL. 

%
%
%


{\bf Human vs.\ Hybrid Judging.} 
%
\textbf{ Table \ref{table:incompletevspredictedqrel}} collects the best AUC performance among the three protocols in each plot of Figure \ref{Figure:infaptaucorrelation}. 
The results from Table \ref{table:incompletevspredictedqrel} suggests that hybrid judging is always superior to human-only judging. Another notable factor is that for high prevalence collections, hybrid judging achieves a $\tau$ correlation of $0.9$ with only using 20\% of human judgments. In contrast, human-only judging requires about 40\% of human judgments to achieve the same $\tau$ correlation (Figure \ref{Figure:infaptaucorrelation}). This establishes the superiority of hybrid labeling both in terms of cost and efficiency since collecting more human judgments is more time consuming and expensive.      

\begin{table}[t]
\small
\begin{center}
\caption{Average (AUC) Kendall's $\tau$ rank correlation achieved, with and without automatic labeling, by the best performing methods in Figure \protect\ref{Figure:infaptaucorrelation} (see its AUC results).} 
\begin{tabular}{|c|c|c|c|c|}
\hline
\bf Labeling & \bf WT'14 & \bf WT'13 & \bf TB'06 & \bf Adhoc'99 \\
\hline
Hybrid    & {\bf 88.8} (SPL) &  {\bf 85.3} (SPL) &  {\bf 84.5} (CAL) & {\bf 88.6} (CAL) \\
Human-only & 81.7 (CAL) &  80.8 (CAL) & 82.3 (CAL) &  88.4 (CAL)\\
\hline
\end{tabular}
\label{table:incompletevspredictedqrel}
\end{center}
\end{table} 

\begin{figure}[h]
\centerline{\includegraphics[width=0.55\textwidth, scale=0.5]{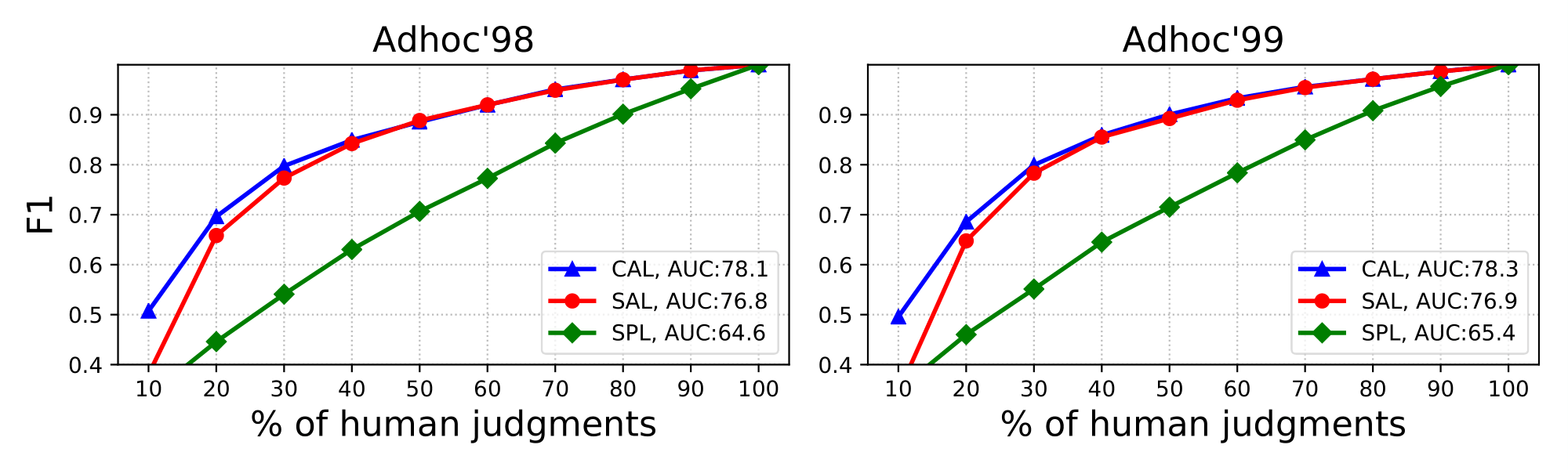}} 
\caption{
Judging cost (x-axis) vs.\ F1 classification accuracy (y-axis) for hybrid human-machine relevance judging for two TREC Tracks. The \% of judgments on x-axis is given wrt.\ the total number of documents in each collection (Table \ref{table:characteristic_data}). Note that ``human judgments'' here encompasses both actual NIST judgments for pool documents and simulated judgments of non-relevance for non-pool documents.
} 
 \label{Figure:f1completecollection}
\end{figure}

\subsection{Evaluation on the Complete Document Collection}

In this section, we evaluate the effectiveness of our method on the  complete document collection. As discussed earlier, whenever the AL system requests a label for a document outside of the pool, we simply return a judgment of non-relevance. 
As in earlier result figures, we plot the percentage of human judgments on the x-axis, but note that: 1) this percentage is now over the entire collection, rather than only the pool; and 2) this percentage encompasses both actual human judgments (from NIST) and simulated judgments for non-pool documents. 
As previously reported, results in this section assume the IS setting. We report evaluations only for Adhoc'98 and Adhoc'99 tracks due to space constraint.

\subsubsection{Classification Results on the Complete Document Collection} {\bf Figure \ref{Figure:f1completecollection}} presents F1 performance results of the three document selection approaches: CAL, SAL
and SPL. As noted earlier, the x-axis of each plot indicates the percentage of documents either judged using  NIST labels (manual) or assumed non-relevant, with the remainder of the collection automatically labeled by the classifier. From Figure \ref{Figure:f1completecollection}, we find that to achieve an F1 measure of 0.9, CAL and SAL need 50\% of hybrid judgments. In contrast, recall that when we evaluated same Adhoc'99 collection assuming the collection was limited to the pool (Figure \ref{Figure:classification}, Bottom Row, IS setting), only 30\% of human judgments were needed to achieve F1 = 0.9. This shows that running AL from scratch over the entire document collection appears more challenging than starting from the filtered pool. One issue, as discussed earlier, is that assuming non-pool documents to always be non-relevant will sometimes be wrong. Given that the pool size is approximately 1/6 of the collection size (for Adhoc'98 and Adhoc'99), AL is roughly 5 times more likely to select a non-pool document for judging which is assumed to be non-relevant. Moreover, as learning progresses, these assumed non-relevant documents will increasingly impact selection of future documents for judging, which could lead to less and less judging of pool documents on which the original TREC evaluation was based.

\subsubsection{Rank Correlation Results on the Complete Document Collection}
\begin{figure}[h]
\centerline{\includegraphics[width=0.55\textwidth, scale=0.5]{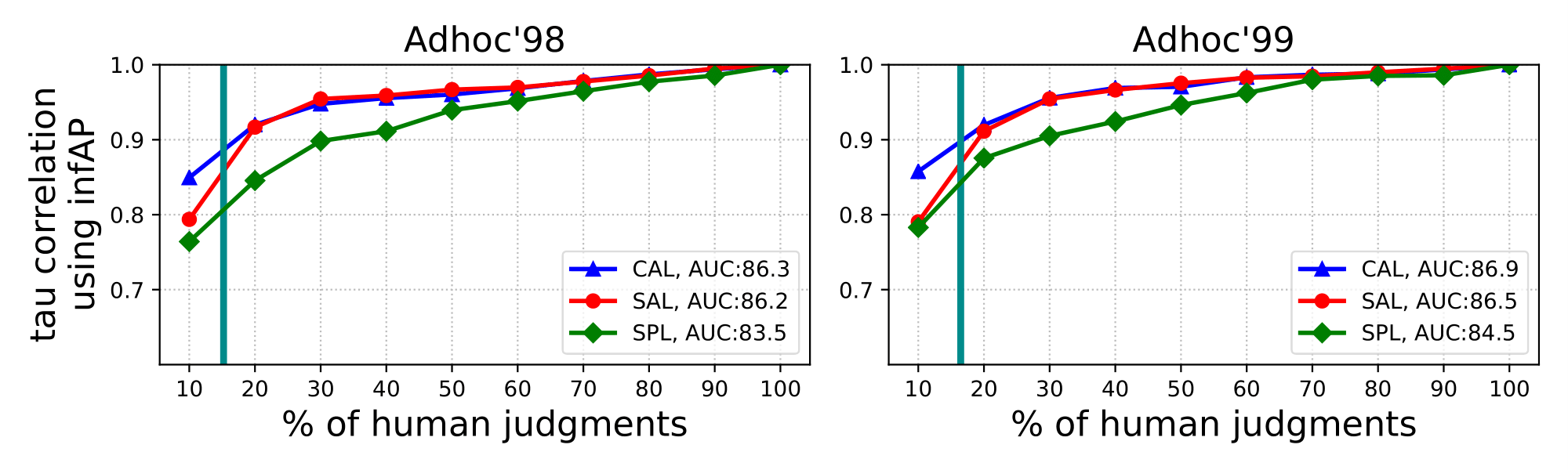}} 
\caption{
Kendall's $\tau$ rank correlation results corresponding to classification results in Figure \ref{Figure:f1completecollection}. Systems are evaluated using infAP. Ground truth ranking is induced by system infAP scores using all NIST judgments. The {\bf dark-cyan colored vertical line} shows the percentage of documents from the complete document collection which were judged by NIST in the original TREC pool. 
} 
\label{Figure:infapcompletecollection}
\end{figure}

We consider how the ranking of systems induced from the hybrid human-classification correlates with the ranking of the same systems induced from NIST judgments ({\bf Figure \ref{Figure:infapcompletecollection}}). From Figure  \ref{Figure:infapcompletecollection}, two (2) observations are immediately evident. Firstly, CAL and SAL consistently outperform SPL in both Adhoc'99 and Adhoc'98 tracks. Secondly, CAL is superior to SAL when the allocated budget is relatively low (e.g.\ when only 10\% of entire document collection are judged, CAL achieves the highest $\tau\approx0.85$, whereas at the same budget SAL achieves only $\tau\approx0.79$ ). Note that these findings are consistent with the experimental evaluation reported on Adhoc'99 collection assuming the collection is limited to the TREC pool only (Figure \ref{Figure:infaptaucorrelation}).     

Finally, we investigate the effectiveness of our AL based approach in constructing a test collection without organizing a shared task in terms of cost. Here the measurement of cost is purely in terms of collected human judgment since quantifying the other costs involved in organizing a shared task 
is beyond the scope of this article. 
Approximately $15.8\%$ of  documents are judged by NIST TREC for these two collections which is the only quantifiable cost associated with NIST TREC along with other  costs. In contrast, our best performing AL approach, CAL, achieves a  $\tau\approx 0.89$ with the original NIST judgments when we  judge the same number of documents ($\approx15.8\%$ which is depicted by dark-cyan colored vertical lines in Figure \ref{Figure:infapcompletecollection}) and unlike TREC, CAL involves no other costs. This shows feasibility of developing a test collection without a shared task. 

\section{CONCLUSION AND FUTURE WORK}
\label{section:conclusion}

We have proposed AL for both intelligent document selection and automatic labeling of unjudged documents. 
Experimental findings suggest that for pool documents, hybrid human-machine labeling approach achieves 90\% F1 in annotating documents when their human counterparts provide on average 58\% fewer annotations (Figure \ref{Figure:classification}). Similarly, in terms of Kendall's $\tau$ rank correlation, hybrid human-machine labeling significantly outperforms the human-only judging across all four test collections (Table \ref{table:incompletevspredictedqrel}). Furthermore, for the complete document collection which contains both pool and non-pool documents, our experimental evidence helps us to hypothesize that assigning a non-relevant label for every non-pool document without considering the actual content of those documents is not beneficial to the document selection and the annotating performance of AL. Despite having these limitations in the complete document collection scenario, by annotating only $\approx15.8\%$ of documents from the document collections, our AL based approach constructs a test collection having a $\tau\approx0.89$ with the original NIST judgments (Figure \ref{Figure:infapcompletecollection}). However, further experiments should be performed on document collections (e.g., WT'14, WT'13, etc.) where the assumption that non-judged documents are non-relevant does not hold completely. 

Various directions remain for future work. Predicting relevance judgments via a classifier introduces the obvious risk of biasing evaluation toward systems applying a similar model for document ranking. How to best handle non-pool documents is also unclear; one could collect new relevance judgments, but they may diverge from relevance criteria of the original judges.

{\bf Acknowledgements.} We thank the reviewers for their valuable feedback. 
This work is supported in part by the Qatar National Research Fund (grant \# NPRP 7-1313-1-245), the Micron Foundation, Wipro, and by {\em Good Systems}\footnote{\url{https://goodsystems.utexas.edu/}}, a UT Austin Grand Challenge to develop responsible AI technologies. The statements made herein are solely the responsibility of the authors.
\bibliographystyle{ACM-Reference-Format}
\bibliography{References}

\end{document}